\documentclass{astlb}

\usepackage{graphicx}
\usepackage{color}

\usepackage{apjfonts}

\usepackage{multirow}
\usepackage[hyperfootnotes=false]{hyperref}

\definecolor{darkblue}{rgb}{0,0,0.9}

\def\smfigure#1#2#3{
  \begin{minipage}{1.0\columnwidth}
    \begin{minipage}{0.049\columnwidth}
      \rotatebox{90}{\small\phantom{0000}#3}
    \end{minipage}
    \begin{minipage}{0.9\columnwidth}
      \includegraphics[bb=40 188 556 678,width=0.97\columnwidth]{#1}
      \centerline{\small #2}
    \end{minipage}

    \vskip 3pt
    ~
  \end{minipage}
}

\begin{document}

\journalinfo{2017}{0}{0}{1}[0]

\title{An extension of the Planck galaxy cluster catalogue}

\author{R. A.~Burenin\email{rodion@hea.iki.rssi.ru}\\
  \vskip -2mm ~\\
  {\rm\em\small  Space Research Institute RAS, Moscow}
}


\shortauthor{R. A. Burenin}

\shorttitle{An extension of the Planck galaxy cluster catalogue}


\submitted{20.11.2016}

\begin{abstract}
  We present a catalogue of galaxy clusters detected in the Planck
  all-sky Compton parameter maps and identified using data from the
  WISE and SDSS surveys. The catalogue comprises about 3000 clusters
  in the SDSS fields. We expect the completeness of this catalogue to
  be high for clusters with masses larger than
  $M_{500}\approx3\times 10^{14}~M_\odot$, located at redshifts
  $z<0.7$. At redshifts above $z\approx0.4$, the catalogue contains
  approximately an order of magnitude more clusters than the 2nd
  Planck Catalogue of Sunyaev-Zeldovich sources in the same fields of
  the sky. This catalogue can be used for identification of massive
  galaxy clusters in future large cluster surveys, such as the
  SRG/eROSITA all-sky X-ray survey.


\end{abstract}

\section{Introduction}

Measurement of the galaxy cluster mass function is one of the most
sensitive methods used to constrain the parameters of the cosmological
model \citep[e.g.,][]{av09a,av09b,PSZcosm13,PSZ2cosm}. For such
studies, large samples of massive galaxy clusters are needed. 

One of the largest samples of massive galaxy clusters is the catalogue
of clusters detected via the Sunyaev-Zeldovich (SZ) effect
\citep{sz72} in the Planck all-sky survey \citep{PSZ1,PSZ2}. In this
survey, the most massive clusters in the observable Universe are
detected nearly uniformly over the entire extragalactic sky. The 2nd
Planck Catalogue of SZ sources \citep[\emph{PSZ2},][]{PSZ2} contains
1653 objects, of which 1203 are confirmed massive galaxy
clusters. Most of these clusters have masses larger than $M_{500} \sim 
6\times 10^{14}~M_\odot$, i.e. they are the most massive clusters in
the Universe. The number density of such objects is very small and
their mass function is very steep.

Since the amplitude of the SZ effect depends mostly on galaxy cluster
mass, lowering the detection limit in the Planck SZ survey should 
enable finding objects of lower mass, which would lead to a rapid  
increase in the number of detected clusters. For example, with a two
times lower detection limit, $M_{500}\sim 3\times 10^{14}~M_\odot$,
the number of detected clusters is expected to increase by an order of
magnitude \citep[e.g.,][]{av09b}. The cluster detection limit could be
lowered if it were possible to use additional data for identification
of clusters and elimination of false detections. 

Below, we demonstrate that a useful sample of galaxy clusters with
masses above $M_{500}\sim 3\times 10^{14}~M_\odot$ at redshifts
$z<0.7$ can be obtained using the Planck all-sky Compton parameter
maps in combination with data from the Wide-Field Infrared Survey 
Explorer (WISE) and Sloan Digital Sky Survey (SDSS). We present a
catalogue of about 3000 galaxy clusters found using these data in the
SDSS fields. 

\section{Source detection in Planck Compton parameter maps}

The detection of SZ sources in the Planck Collaboration catalogues was 
done using specialized procedures that take the spectral
and spatial shape of the source into account \citep[see, e.g.,][]{PSZ2}. In
addition, Compton parameter maps ($y$-maps) were constructed
\citep{PYmap}, which were mainly used for studying the angular power
spectrum of the SZ signal. It has been demonstrated that there is a
good agreement between the objects from the SZ source catalogue and
the sources detected in the $y$-maps. Therefore, for simplicity, we
have used the Planck Compton parameter maps to detect galaxy clusters. 

We exploited maps of the Compton parameter and its standard deviation from
the Planck 2015 data release \citep{PYmap}, as provided by the Planck
Legacy
Archive\footnote{\href{http://pla.esac.esa.int/}{http://pla.esac.esa.int/}}. We
performed our source search using the NILC maps, because they have
somewhat lower noise at small angular scales \citep[see details in][]{PYmap}. We
smoothed the standard deviation map with a $1^\circ$-radius median
filter. We then obtained a signal to noise map from the $y$-map and
the smoothed standard deviation map and additionally subtracted large
scale anisotropy, which was estimated by smoothing the map with a
$1^\circ$-radius median filter. 

As a Galaxy foreground mask we used a mask produced by
\cite{khatri16}. This mask makes allowance not only for Galaxy dust
emission but also for the emission of carbon monoxide (CO), in
particular in high-latitudes molecular clouds. The spectrum of the CO
signal resembles the spectrum of $y$-distortions, so that CO emission
significantly contaminates the Planck $y$-maps. Specifically, we used
the 61\% CO mask available in the public domain\footnote{\href{http://theory.tifr.res.in/~khatri/szresults/}{http://theory.tifr.res.in/${}_{\textrm{\symbol{126}}}$khatri/szresults/}}. 

\begin{figure}
  \centering \smfigure{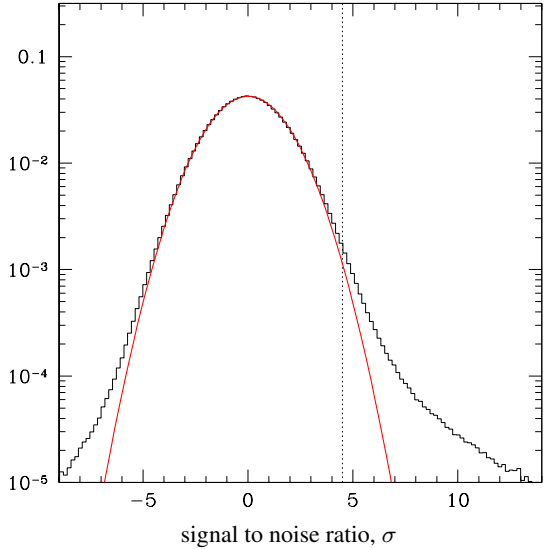}{signal to noise ratio,
    $\sigma$}{}
  \caption{Signal to noise ratio distribution of the $y$-parameter inside
    the Galactic foreground mask. The source detection threshold is shown
    with the dotted line.}
  \label{fig:dist}
\end{figure}

Figure~\ref{fig:dist} shows the $y$-parameter signal to noise ratio
distribution obtained as discussed above inside the Galactic
foreground mask. Small deviations of the $y$-parameter from the mean value
are well described by a Gaussian distribution, since the observed signal
consists of instrumental noise and the averaged signal of a large number of
faint SZ sources. At signal to noise ratios approximately $>5$, the
distribution differs significantly from the Gaussian one. Obviously,
individual SZ sources start to appear above the noise level at
these deviations from the mean value.

Since we use additional IR and optical data for cluster
identification, which allows us to effectively eliminate false
detections from the sample, we can use a lower source detection
threshold to detect sources in the $y$-parameter map. We decided to
use a $4.5\sigma$ source detection threshold (shown with the dotted
line in Fig.~\ref{fig:dist}). Although a lot of projections of
faint SZ sources may appear near this threshold, we expect to find
also many real individual massive clusters, which can be identified
using IR and optical data. 

We considered only SZ sources inside the foreground mask discussed
above. Also, only sources at Galactic latitudes $|b|>20^\circ$ were
considered, otherwise source identification in optical and IR bands
would be difficult due to strong contamination from Galactic stars. We
have thus selected 20290 SZ sources, of which 9227 are located in the
SDSS fields. This number of detected sources indicates that the cluster
mass threshold in our sample has been lowered not more than two
times compared to the 2nd Planck Catalogue of SZ sources. Therefore,
these sources (apart from the nearest ones) are not expected to be
identified with clusters less massive than $M_{500}\sim 3\times
10^{14}~M_\odot$. 

\section{Identification of clusters in the optical and infrared}

\subsection{The \emph{redMaPPer} catalogue}

Galaxy clusters can be efficiently found using optical photometry,
since most galaxies in clusters have similar colours 
and form the so-called red sequence in the colour-magnitude diagram
\citep[e.g.,][]{gladdersyee00}. To identify SZ sources detected in the 
Planck Compton parameter map, we used a galaxy cluster catalogue obtained
from SDSS data using the \emph{redMaPPer} cluster detection
algorithm. Specifically, we used the publicly available \emph{redMaPPer}
catalogue, version
6.3\footnote{\href{http://risa.stanford.edu/redMaPPer/}{http://risa.stanford.edu/redMaPPer/}},
which contains about 26000 galaxy clusters, all having relatively
good photometric redshift and richness estimates. Cluster richness is
known to correlate well with total cluster mass.

Since we are interested in massive galaxy clusters, with masses above
$M_{500}\sim 3\times 10^{14}~M_\odot$, we considered only clusters
with $\lambda>40$, where $\lambda$ is the cluster richness provided in
the \emph{redMaPPer} catalogue. This restriction allows us to reject
low mass clusters but to keep nearly all clusters with masses
$M_{500}>3\times 10^{14}~M_\odot$ included in the reference sample
\citep{RMPlanck15}.

\begin{figure}
  \centering \smfigure{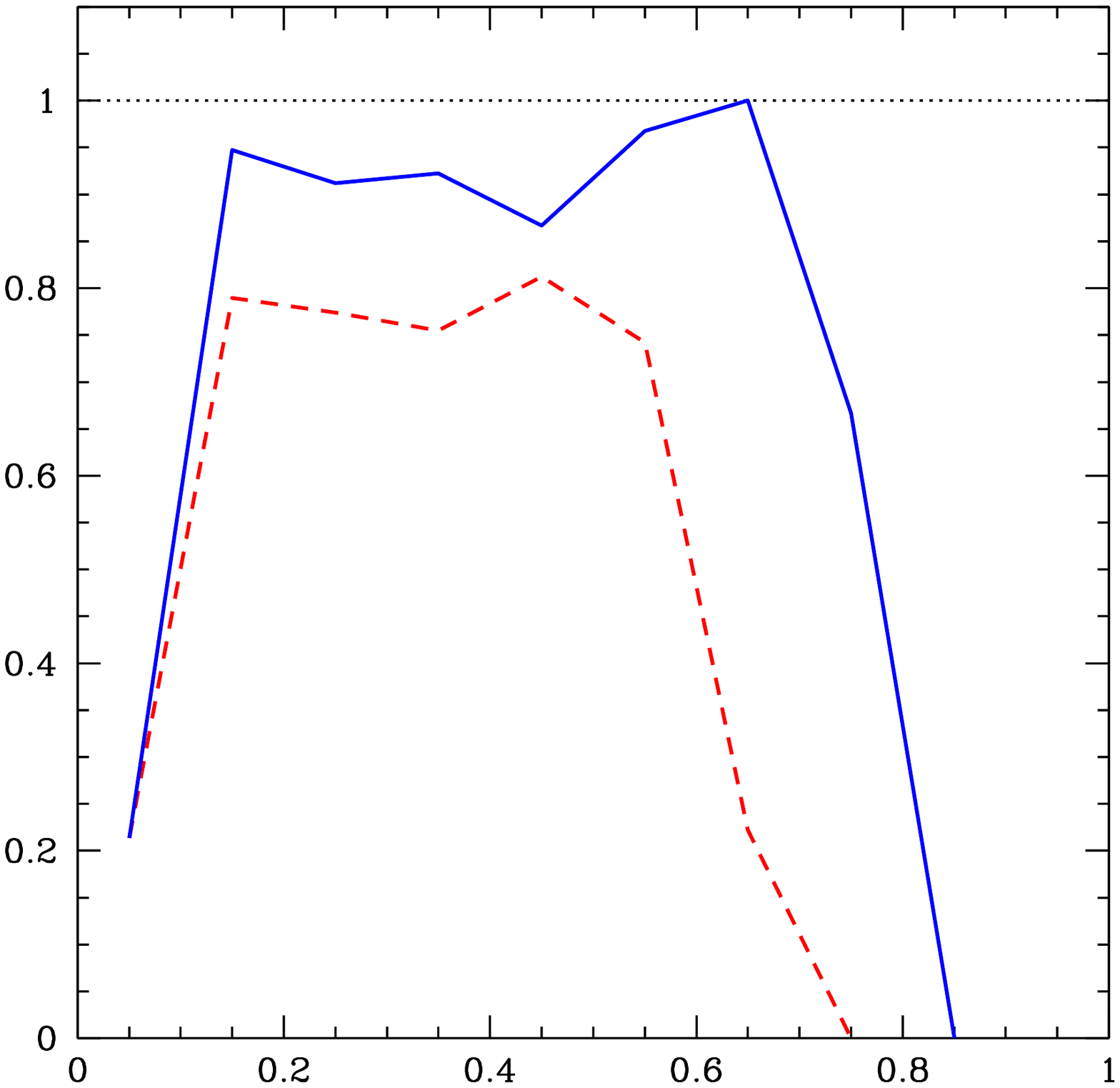}{$z$}{fraction of identified
    clusters}
  \caption{Fraction of confirmed clusters from the PSZ2 catalogue in
    the SDSS fields that have also been detected with the \emph{redMaPPer}
    algorithm (red dashed line, the \emph{redMaPPer} mask is not taken
    into account) and with our new algorithm using WISE IR data
    (solid blue line).} 
  \label{fig:detfrac}
\end{figure}

Figure~\ref{fig:detfrac} shows (red dashed line) the fraction of
confirmed clusters from the PSZ2 catalogue that are detected in the
SDSS fields using the \emph{redMaPPer} algorithm (according to the
version of the catalogue used here). We have applied the
richness limit discussed above, but this has almost no effect on the
number of identified clusters, as expected. Note that the
\emph{redMaPPer} mask has not been taken into account here, which may
be one of the main reasons why the maximum identified fraction is less
than 100\%. We also see that the fraction of detected clusters
drops at redshifts $z>0.55$.

\subsection{Clusters identified using WISE and SDSS data}

In order to more efficiently identify clusters at higher redshifts, we
used IR data from the WISE survey in addition to SDSS optical data. The
WISE all-sky survey \citep{wright10} started in 2009 and was
initially done in four photometric bands: 3.4, 4.6, 12 and
22~$\mu$m. Since the end of the cryogenic phase in 2010, the survey 
has been continuing in the 3.4 and 4.6~$\mu$m bands
\citep{mainzer14}. For galaxy cluster observations, the 3.4~$\mu$m
photometric band is most useful. In this band, distant 
galaxy clusters are well detected at redshifts up to $z\approx 1$--$2$
\citep[e.g.,][]{br15}.

To search for galaxy clusters in the 3.4~$\mu$m band images of the WISE
all-sky survey, we used a completely automated algorithm that builds
on the procedure described in our previous paper \citep{br15}. To
detect clusters, we first subtracted stars from the WISE images, then
detected extended IR sources in these images by convolving them 
with $\beta$-models of various angular sizes, and finally identified
the brightest cluster galaxies and red sequences inside the detected
IR sources using SDSS photometric data. 

As compared to our earlier work cited above, the following
improvements were made. We used more recent coadds of WISE and NEOWISE
images, presented in \cite{meisner16} and available for public
use\footnote{\href{http://unwise.me/}{http://unwise.me/}} \citep[see
also, ][]{lang14}. Flux measurements for the sources in WISE images
were made using a more complete PSF model, taking into account not only
its wings at large angular scales but also its angular asymmetry
relative to its center. The data on source positions were
taken from SDSS, data release 13 \citep{sdssdr13}. Source
flux fitting was done with frozen source positions in the sky,
i.e.\ using so-called ``forced photometry''. For brighter galaxies,
where the form of the galaxy should be taken in account in addition to
the PSF model, we used the ``forced photometry'' from \cite{lang16}. In
addition, we improved the red sequence detection procedure. The
current version of this cluster detection algorithm is a preliminary
one. We plan to further improve it in the future, but even this
preliminary version is suitable for identification of massive galaxy
clusters among Planck SZ sources.

\begin{figure}
  \centering
  \smfigure{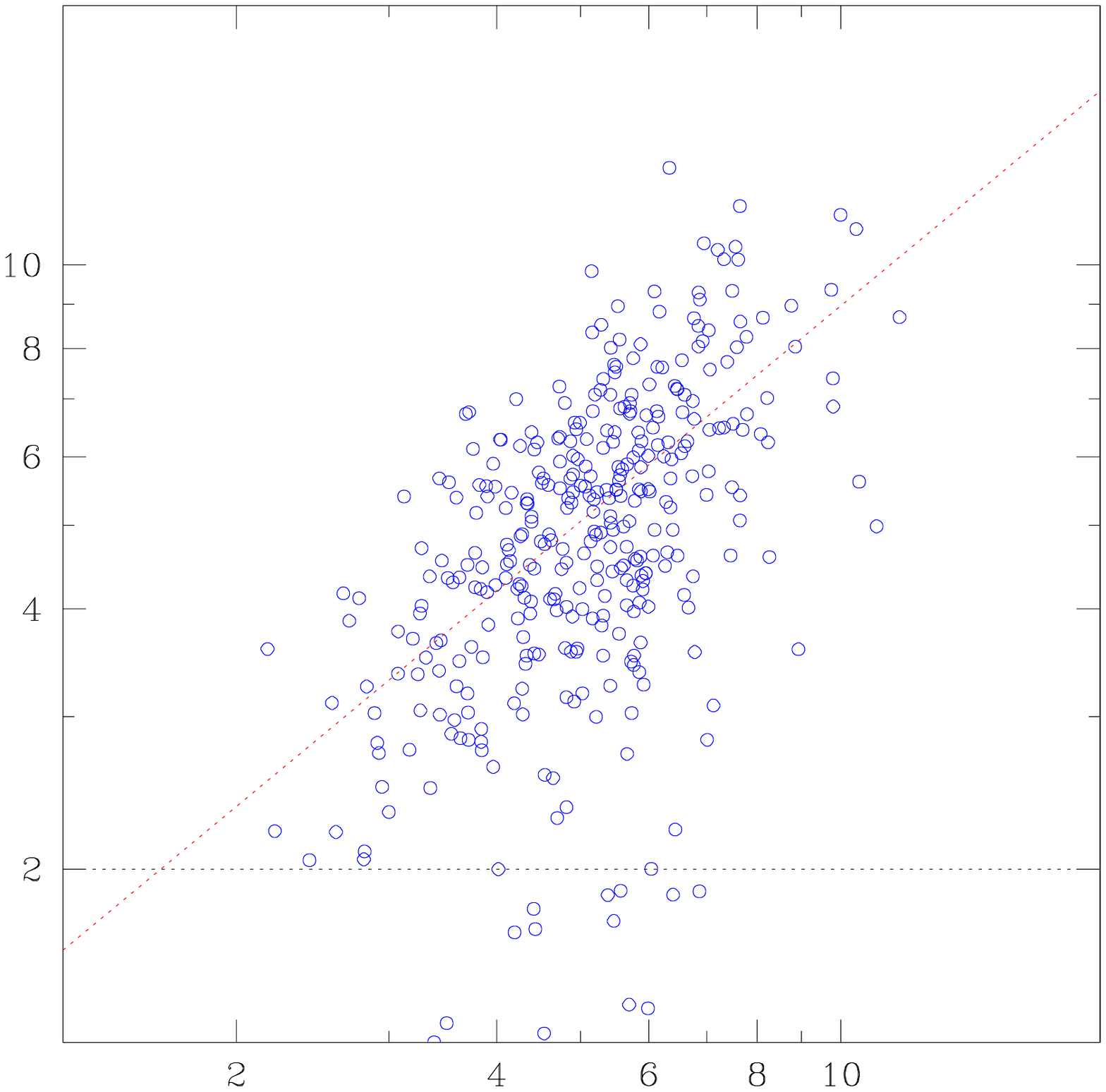}{$M_{500},
    10^{14}~M_\odot$}{$L_{3.4\mbox{\scriptsize $\mu$m}}, 10^{44}$~erg/s}
  \caption{Relation between the 3.4~$\mu$m band luminosity
    inside the 1~Mpc radius and the total gravitational mass $M_{500}$
    of clusters from the PSZ2 catalogue.}
  \label{fig:mszwlum}
\end{figure}

\begin{figure}
  \centering
  \smfigure{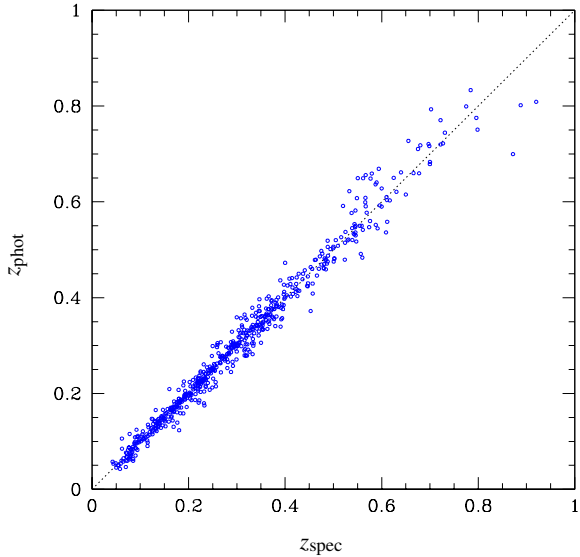}{$z_{\mbox{\scriptsize spec}}$}{$z_{\mbox{\scriptsize phot}}$}
  \caption{Spectroscopic galaxy cluster redshifts vs.\ their
    photometric estimates from SDSS and WISE photometric data.}
  \label{fig:zphot}
\end{figure}

Since only massive clusters can appear among the SZ sources from the
Planck survey, we should look for clusters with masses above
$M_{500}\sim 3\times 10^{14}~M_\odot$ in WISE images. In order to
estimate galaxy cluster masses from IR data, cluster IR luminosities
can be used
\citep[e.g.,][]{lin04,kopylova06}. Figure~\ref{fig:mszwlum} shows the
correlation between the IR luminosity\footnote{Hereafter, the
  cosmological model with $\Omega_m=0.3$, $\Omega_\Lambda=0.7$ and
  $H_0=70$~km~с$^{-1}$~Mpc$^{-1}$ is adopted.} in the 3.4~$\mu$m band
and the total mass $M_{500}$ for clusters in the PSZ2 catalogue. The
IR luminosity was calculated as the combined luminosity of red
sequence galaxies within the 1~Mpc projected radius. The zero point
and bandwidth calibrations were adopted from \cite{jarrett11}. The
K-corrections in the 3.4~$\mu$m band were calculated using the 11~Gyr
age synthetic stellar population model template taken from
\cite{bc03}, which appears to be a suitable model for all red sequence
colours for clusters at redshifts $z<0.8$.

We see from Fig.~\ref{fig:mszwlum} that the cluster IR luminosity
estimated in this way correlates well with the total cluster mass. The
power law slope of this correlation is $\approx 0.83\pm0.05$. The
slope is smaller than unity since the luminosity is calculated within
a constant physical radius, rather than within a radius of constant
density contrast. The scatter of the IR luminosity near this
correlation, excluding larger than $2.5\sigma$ deviations, is
$\sigma_{\ln L}= 0.25$ or $\pm29\%$. Therefore, IR luminosities allow
us to obtain total cluster mass estimates that are nearly as accurate
as mass estimates obtained from cluster X-ray luminosities
\citep{av09b} or optical richness \citep{RMPlanck15}.

To identify clusters among Planck SZ sources, we used only clusters
with IR luminosities $L_{3.4\mbox{\scriptsize $\mu$m}}>2\times
10^{44}$~erg~s$^{-1}$. We see from Fig.~\ref{fig:mszwlum} that most of
the clusters with masses $M_{500}> 3\times 10^{14}~M_\odot$ have IR
luminosities above this limit. The fraction of confirmed clusters from
the PSZ2 catalogue in the SDSS fields that are detected with our
automated procedure, with the above IR luminosity constraint applied, 
is shown in Fig.~\ref{fig:detfrac} by the solid blue line. We see
that nearly 90\% of PSZ2 clusters at $z>0.1$ are detected using WISE
data, while this fraction drops rapidly at $z>0.7$. Therefore, we have
managed to significantly increase the completeness of cluster
identification at redshifts $0.5<z<0.7$, as compared to the
\emph{redMaPPer} catalogue.

Figure~\ref{fig:zphot} compares photometric redshift estimates
obtained with our automated cluster detection procedure with
spectroscopic redshifts. These estimates were obtained using the
cluster red sequence colours and the magnitudes of cluster brightest
galaxies from SDSS and WISE photometric data. To calibrate our
photometric estimates, we used clusters from the PSZ2 catalogue
\citep{PSZ2} as well as from the 400 square degree X-ray cluster survey
\citep[\emph{400d},][]{400d} and the earlier 160 square degree cluster
survey \citep[\emph{160d},][]{160d} based on ROSAT pointings data. To
increase the number of high redshift clusters at $z>0.6$, five
additional clusters were taken from the EMSS \citep{gioia94}, WARPS
\citep{warps1,warps2}, MACS \citep{macs} and NEP \citep{henry06}
surveys, which are compiled in the MCXC catalogue \citep{mcxc}. 

The accuracy of our photometric redshift estimation is
$\delta{z_{\mbox{\scriptsize phot}}}/(1+z) \approx 0.01$ at $z<0.5$
and $\delta {z_{\mbox{\scriptsize phot}}}/(1+z) \approx 0.03$ at
$0.5<z<0.7$. We see from Fig.~\ref{fig:zphot} that our photometric
redshift estimates, and therefore cluster IR luminosities, are not
reliable at $z>0.7$. Therefore, to identify Planck SZ sources, we use
below only clusters with photometric redshift estimates
$z_{\mbox{\scriptsize phot}}<0.7$.

\begin{table*}
  \renewcommand{\arraystretch}{1.2}
  \renewcommand{\tabcolsep}{6pt}
  \centering
  \caption{The cluster catalogue}
  \label{tab:cat}
  \medskip
  \footnotesize
  \begin{tabular}{lccrrrlll}
    \hline
    \hline
    Number & $\alpha$ & $\delta$ & S/N & $\delta\alpha$~ & $\delta\delta$~ & ~$z$ & $N_z$ & Note \\
    & \multicolumn{2}{c}{(J2000)} & & \multicolumn{2}{c}{~~arc min}& \\
    \noalign{\vskip 3pt\hrule\vskip 3pt}
    
    1 & $00~00~48.9$ & $+29~06~52$ & 4.58 & $ 0.37$ &$ 3.64$ & $0.1928$ & $3$ & \scriptsize  \\
2 & $00~01~11.8$ & $+21~33~49$ & 7.77 & $-0.42$ &$ 2.01$ & $0.4110$ & $5$ & \scriptsize PSZ2\,G107.67-39.78 \\
3 & $00~01~24.0$ & $-00~00~53$ & 4.55 & $-0.43$ &$ 0.97$ & $0.2479$ & $5$ & \scriptsize  \\
4 & $00~02~01.1$ & $+12~03~28$ & 9.52 & $ 0.62$ &$-0.44$ & $0.1989$ & $17$ & \scriptsize PSZ2\,G104.30-48.99 \\
5 & $00~02~49.6$ & $-01~04~06$ & 4.99 & $-4.60$ &$ 4.99$ & $0.7600$ & $1$ & \scriptsize $^{**}$   \\
6 & $00~02~47.9$ & $-05~48~37$ & 5.86 & $-3.21$ &$ 4.33$ & $0.4586$ & $1$ & \scriptsize  \\
7 & $00~03~06.0$ & $-06~05~13$ & 14.20 & $-1.48$ &$ 0.47$ & $0.2334$ & $3$ & \scriptsize $^{**}$  PSZ2\,G092.16-66.01 \\
8 & $00~03~33.3$ & $+10~02~17$ & 6.10 & $-2.59$ &$ 0.83$ & $0.3707$ & $4$ & \scriptsize  \\
9 & $00~03~45.3$ & $+02~04~55$ & 7.21 & $-1.09$ &$ 0.93$ & $0.0924$ &  & \scriptsize $^{**}$  PSZ2\,G099.57-58.64 \\
10 & $00~04~01.1$ & $+30~41~26$ & 5.52 & $ 1.43$ &$-0.96$ & $0.7397$ & $2$ & \scriptsize  \\
11 & $00~03~58.1$ & $-11~00~51$ & 4.87 & $ 0.21$ &$ 3.00$ & $0.241^{*}$ &  & \scriptsize  \\
12 & $00~04~09.2$ & $+04~35~54$ & 5.36 & $-1.55$ &$ 0.21$ & $0.6396$ & $1$ & \scriptsize  \\
13 & $00~05~24.6$ & $+16~11~23$ & 7.22 & $-0.11$ &$-1.81$ & $0.1155$ & $13$ & \scriptsize  \\
14 & $00~06~16.1$ & $-10~21~40$ & 5.19 & $ 1.57$ &$ 4.39$ & $0.2180$ & $1$ & \scriptsize $^{**}$   \\
15 & $00~06~21.4$ & $+10~53~57$ & 11.20 & $ 0.26$ &$ 2.14$ & $0.1669$ & $15$ & \scriptsize PSZ2\,G105.40-50.43 \\
16 & $00~07~01.8$ & $+25~05~02$ & 5.64 & $ 1.36$ &$-1.29$ & $0.2409$ & $2$ & \scriptsize  \\
17 & $00~07~06.0$ & $+10~34~21$ & 4.83 & $-0.17$ &$-0.26$ & $0.1649$ & $6$ & \scriptsize  \\
18 & $00~07~27.4$ & $+12~33~49$ & 6.80 & $ 2.02$ &$-3.61$ & $0.697^{*}$ &  & \scriptsize  \\
19 & $00~08~12.4$ & $+02~03~14$ & 8.91 & $ 0.42$ &$ 1.98$ & $0.3651$ & $3$ & \scriptsize PSZ2\,G101.55-59.03 \\
20 & $00~08~57.1$ & $+13~06~46$ & 4.68 & $-0.55$ &$-0.26$ & $0.195^{*}$ &  & \scriptsize $^{**}$   \\
21 & $00~09~01.8$ & $+32~10~09$ & 5.42 & $ 2.26$ &$-2.13$ & $0.4790$ & $2$ & \scriptsize  \\
22 & $00~09~13.6$ & $+03~56~18$ & 4.83 & $-3.27$ &$-3.06$ & $0.1015$ & $2$ & \scriptsize $^{**}$   \\
23 & $00~09~20.2$ & $+06~49~06$ & 10.04 & $-0.09$ &$-0.34$ & $0.2361$ & $2$ & \scriptsize PSZ2\,G104.71-54.54 \\
24 & $00~09~42.4$ & $+03~43~36$ & 4.72 & $ 1.23$ &$ 1.60$ & $0.5622$ & $2$ & \scriptsize  \\
25 & $00~09~54.6$ & $+12~17~28$ & 8.18 & $-3.89$ &$-0.63$ & $0.1745$ & $10$ & \scriptsize  \\
26 & $00~09~59.7$ & $+17~54~37$ & 7.46 & $ 3.71$ &$ 0.08$ & $0.5583$ & $2$ & \scriptsize  \\
27 & $00~10~03.9$ & $+25~52~01$ & 7.46 & $ 0.08$ &$-0.98$ & $0.3211$ & $6$ & \scriptsize $^{**}$   \\
28 & $00~10~08.0$ & $+33~06~23$ & 5.65 & $-0.15$ &$-0.84$ & $0.1136$ & $1$ & \scriptsize $^{**}$   \\
29 & $00~10~13.5$ & $+17~44~40$ & 8.31 & $-1.42$ &$ 0.47$ & $0.1715$ & $5$ & \scriptsize PSZ2\,G109.22-44.01 \\
30 & $00~10~18.2$ & $+06~39~47$ & 8.59 & $ 0.48$ &$-0.75$ & $0.2648$ & $3$ & \scriptsize PSZ2\,G104.98-54.79 \\
31 & $00~10~26.6$ & $+11~31~16$ & 6.37 & $-2.58$ &$ 1.32$ & $0.0924$ & $4$ & \scriptsize  \\
32 & $00~10~48.4$ & $+29~10~23$ & 9.68 & $-1.09$ &$ 0.50$ & $0.3332$ & $6$ & \scriptsize PSZ2\,G112.35-32.86 \\
33 & $00~10~50.8$ & $-01~02~34$ & 5.83 & $-3.54$ &$-1.27$ & $0.4646$ & $1$ & \scriptsize $^{**}$   \\
34 & $00~10~56.0$ & $+18~29~42$ & 4.65 & $ 0.74$ &$-2.00$ & $0.5668$ & $2$ & \scriptsize  \\
35 & $00~11~45.5$ & $+32~24~27$ & 16.12 & $ 0.29$ &$-1.04$ & $0.1012$ & $15$ & \scriptsize PSZ2\,G113.29-29.69 \\
36 & $00~12~13.5$ & $+14~00~41$ & 9.62 & $-0.39$ &$-0.61$ & $0.3895$ & $2$ & \scriptsize PSZ2\,G108.71-47.75 \\
37 & $00~12~33.4$ & $-00~16~23$ & 5.07 & $ 0.86$ &$-2.66$ & $0.4027$ & $2$ & \scriptsize  \\
38 & $00~12~43.8$ & $+06~08~08$ & 5.26 & $-0.20$ &$ 4.55$ & $0.4310$ & $1$ & \scriptsize  \\
39 & $00~12~47.7$ & $-08~58~27$ & 8.86 & $-0.37$ &$-2.76$ & $0.3384$ & $3$ & \scriptsize $^{**}$  PSZ2\,G094.46-69.65 \\
40 & $00~14~58.3$ & $-00~55~06$ & 6.03 & $ 1.14$ &$ 2.08$ & $0.5349$ & $5$ & \scriptsize  \\

    \dots & \dots & \dots & \dots~~~ & \dots & \dots & \dots\\
    \noalign{\vskip 2pt\hrule\vskip 2pt}
  \end{tabular}
  
  \medskip
  \begin{minipage}{0.87\linewidth}
    \footnotesize
     
    $^{*}$ Photometric redshift estimate.
    \medskip
    
    $^{**}$ Two or more clusters at different redshifts are found in
    the SZ source field. The SZ source is identified with the cluster
    with the largest mass estimate from SDSS and WISE data. \medskip
    
    Note. Only a small part of the Table is presented here for
    reference. The complete table contains 2964 lines and is available
    in the electronic version of the journal and also at
    \href{http://hea.iki.rssi.ru/psz/en/}{http://hea.iki.rssi.ru/psz/en/}
    
  \end{minipage}

\end{table*}

\section{The cluster catalogue}

As discussed above, there are 9227 SZ sources in the Planck
Compton parameter map in the SDSS fields at $|b|>20$, inside our
Galaxy foreground map. We cross-correlated this sample with the
following objects: 

\begin{itemize}
\item objects from the 2nd Planck Catalogue of SZ sources
  \citep[\emph{PSZ2},][]{PSZ2}; 

\item massive galaxy clusters from the \emph{redMaPPer} catalogue
  \citep{redmapper14};
  
\item massive galaxy clusters detected in the fields of the SZ
  sources by our automated procedure using WISE and SDSS data.
\end{itemize}

As a result, we identified 2964 massive galaxy clusters with Planck SZ
sources. Out of them, 483 clusters are present in the PSZ2 
catalogue (see below), 1795 were identified using the
\emph{redMaPPer} catalogue and 2573 using WISE and SDSS data, as
discussed above. Many clusters are identified with several methods. For
example, 1459 clusters are identified both with the \emph{redMaPPer}
catalogue and using WISE and SDSS data. 

\begin{figure}
  \centering \smfigure{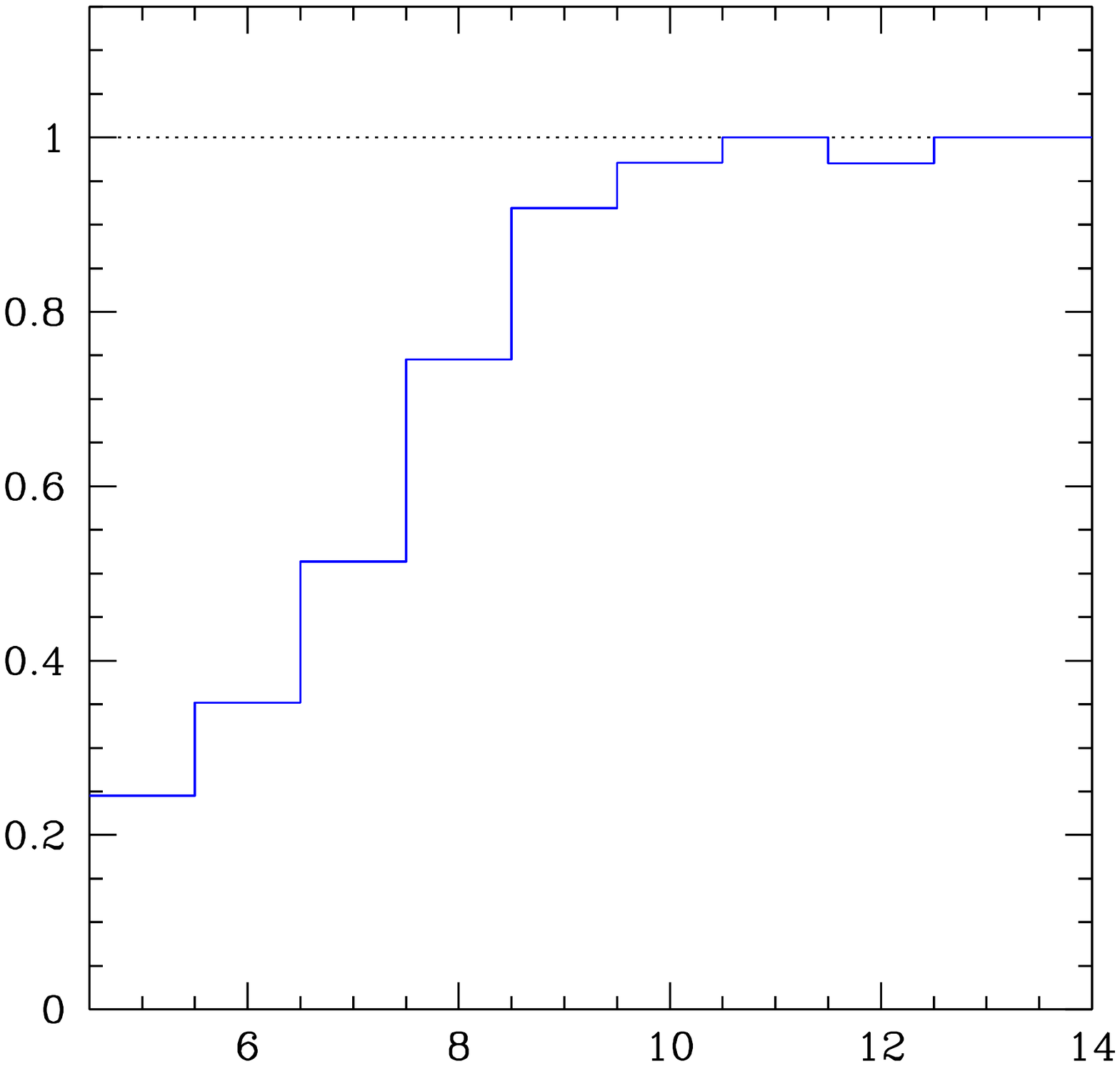}{signal to noise,
    $\sigma$}{fraction of identified clusters}
  \caption{Fraction of SZ sources identified with massive galaxy
    clusters as a function of the signal to noise ratio in the Planck
    $y$-parameter map.} 
  \label{fig:detfrac_all}
\end{figure}

Figure~\ref{fig:detfrac_all} shows the fraction of SZ sources identified
with massive galaxy clusters as a function of the signal to noise
ratio in the Planck $y$-parameter map. At low signal to noise ratios, 
only a small fraction of sources are identified with massive galaxy
clusters. The other SZ sources are probably associated with
non-gaussianities of the $y$-parameter maps, which are expected to
arise due to projections of less massive clusters and groups of galaxies.

The catalogue of identified clusters is presented in
Table~\ref{tab:cat}, which provides the number of the source in the
list, its coordinates ($\alpha$, $\delta$, J2000), significance in the
$y$-parameter map (S/N), the shift of the optical center relative to
the center of the SZ source ($\delta\alpha$, $\delta\delta$ in
arcminutes), cluster redshift ($z$) and the number of galaxies used to
obtain this redshift ($N_z$, see below for details). The last column
of the Table gives the name of the source in the PSZ2 catalogue and
indicates ambiguous identifications, in which case we identified the
SZ source with the most massive cluster, as estimated from optical and
IR data. The number of such projections in our sample is large, about
37\%. Note that the fraction of projections for Planck SZ sources is
relatively high even for bright SZ sources due the insufficient
angular resolution of the Planck telescope
\citep[][]{pipIV_13,planckRTT150}.

\subsection{The cluster redshifts}

The cluster redshifts given in Table~\ref{tab:cat} in most cases are
obtained from SDSS spectroscopic redshifts of galaxies. This is
possible because brightest cluster galaxies are included in the
samples of luminous red galaxies (LRG) that are observed
spectroscopically in SDSS for studying baryon acoustic oscillations
\citep{sdss-lrg}. In addition, spectroscopy of cluster member galaxies
for X-ray selected clusters \citep{sdss-spiders} has recently been
started within SDSS.

\begin{figure}
  \centering \smfigure{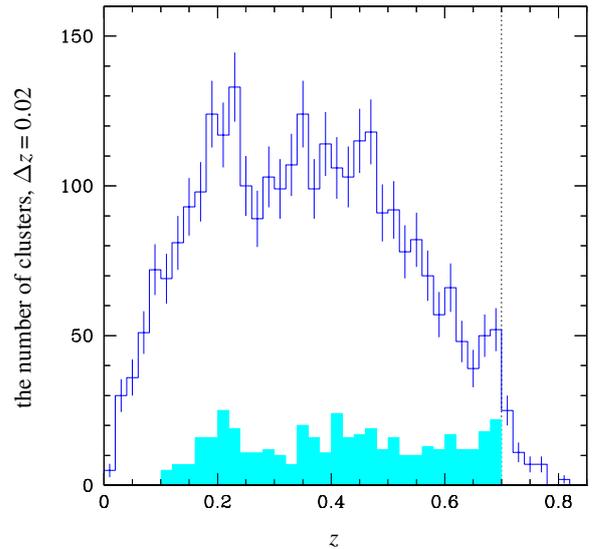}{$z$}{the number of clusters, $\Delta z=0.02$}
  \caption{Redshift distribution of identified clusters. The grey
    filled histogram shows the distribution of clusters without
    spectroscopic redshifts.} 
  \label{fig:dsz}
\end{figure}

For the SDSS based spectroscopic redshifts, Table~\ref{tab:cat}
further provides the number of galaxies that were used to obtain the
redshift of the cluster ($N_z$). For these measurements, we selected
galaxies at $<540$~kpc projected angular distances from the optical
center of the cluster, whose spectroscopic redshifts are within the
photometric redshift estimate uncertainty for the cluster. We then
found the cluster spectroscopic redshift as the median value of the
redshifts of these galaxies, having rejected $\delta z>0.01$ redshift
deviations from the median. If the dispersion of the spectroscopic
redshifts was too high and the median redshift value could not be
obtained in this way, we adopted the photometric redshift
estimate. Note that caution should be given to spectroscopic redshifts
based on a small number of galaxies spectra, especially if there is no
spectroscopic redshift for the brightest cluster galaxy. However, in
most cases the accuracy and reliability of the spectroscopic redshifts
is much higher then for the photometric redshift estimates.

It turns out that spectroscopic redshifts can be obtained from SDSS
data for the majority of clusters in our catalogue, namely for 2541
clusters out of 2964. Figure~\ref{fig:dsz} shows the redshift 
distribution of the clusters from the catalogue. There are 423
clusters without spectroscopic redshifts in the catalogue. Their
redshift distribution is shown in Fig.~\ref{fig:dsz} with the grey filled
histogram.

\subsection{Clusters from PSZ2 catalogue}

There are 1174 objects in the 2nd Planck Catalogue of
Sunyaev-Zeldovich sources \citep[\emph{PSZ2},][]{PSZ2} at $|b|>20$,
inside the Galaxy foreground mask. Out of them, we detected 1135
objects in the Planck Compton parameter maps. Most of these sources
are detected with a signal to noise ratio $>7$.

We did not detect 39 objects from the PSZ2 catalogue in the Planck
$y$-parameter map. Among these, \emph{PSZ2\,G006.84+50.69},
\emph{PSZ2\,G056.62$+$88.42}, \emph{PSZ2\,G061.75$+$88.11} and
\emph{PSZ2\,G341.09$-$33.15} are located at the edges of very bright
SZ sources --- massive galaxy clusters A2029, Coma and A3667. The
pair of sources \emph{PSZ2\,G096.77$-$50.29} and
\emph{PSZ2\,G096.78$-$50.20} is located at an angular distance
$5.5\arcmin$ and is thus a duplicate.

We could not identify the reasons why the other PSZ2 sources are not
detected in the $y$-parameter map. We note however that there are only
three confirmed clusters among these sources. Moreover, the optical 
images of most of these sources reveal molecular clouds, which could
produce a false SZ signal due to CO emission
\citep{khatri16}. Therefore, almost all of the PSZ2 sources not
included in our SZ source list are false ones or not confirmed as
galaxy clusters.

There are 519 sources from the PSZ2 catalogue in the SDSS fields, 
inside our mask. Among them, 483 are identified with galaxy clusters in the 
optical and IR, as discussed above. Out of the remaining 36 sources,
two sources (\emph{PSZ2\,G152.47$+$42.11} and
\emph{PSZ2\,G199.73$+$36.98}) could be identified with galaxy clusters
with optical richness and IR luminosity lower than our
thresholds. In addition, a few more clusters may be identified with
distant clusters, located at $z>0.7$. Currently we have only
photometric redshift estimates for these clusters, and they are thus
not included in our catalogue. The other SZ sources from this list
cannot be confirmed as galaxy clusters using the available data.

Thus, about 93\% of the sources from the PSZ2 catalogue that are also
included in our SZ source list are confirmed as real galaxy
clusters. This fraction is notably higher than the fraction
of confirmed clusters in the PSZ2 catalogue (1203 out of 1653, i.e.\ about
73\%). The reason is that we used a CO emission mask to exclude false
detections \citep{khatri16} and also that we confirmed 28 new
clusters from the PSZ2 catalogue (see Table~\ref{tab:cat}).

\section{Discussion}

We have found 2964 galaxy clusters in the SDSS fields at $|b|>20$,
inside the Galaxy foreground mask, while there are only 483 clusters 
in the 2nd Planck Catalogue of Sunyaev-Zeldovich sources
\citep[][]{PSZ2} in the same area of the sky. The difference between
the number of clusters in our list and in the PSZ2 catalogue is even
higher for distant clusters. For example, there are 1218 clusters at
$z>0.4$ in our catalogue, while there are only 97 clusters in the PSZ2
catalogue at these redshifts in the same regions of the sky. As
discussed above, this is in agreement with the estimates of the lower mass
limit for our catalogue, $M_{500} \sim 3\times 10^{14}~M_\odot$
at high redshifts, which is approximately two times lower than for the
PSZ2 catalogue.

\begin{figure}
  \centering \smfigure{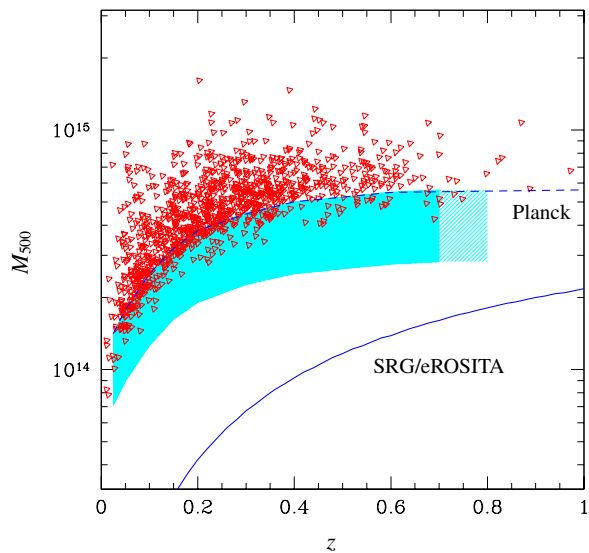}{$z$}{$M_{500}$}
  \footnotesize
  \vskip -4.9cm \hskip 5.7cm Planck
  \vskip 4.6cm 
  \vskip -2.9cm \hskip 3cm SRG/eROSITA
  \vskip 2.5cm 

  \caption{The shaded area shows the approximate range of masses and
    redshifts for clusters from our catalogue. Red points show masses
    and redshifts for clusters from the PSZ2 catalogue \citep{PSZ2}. Also
    approximate lower limits for masses of clusters from the PSZ2
    catalogue and the future SRG/eROSITA survey are shown with the
    dashed and solid lines, respectively.}
  \label{fig:m_z}
\end{figure}

Figure~\ref{fig:m_z} shows an approximate range of masses and
redshifts for clusters from our catalogue. For comparison, we show 
the masses and redshifts for clusters from the PSZ2 catalogue and
lower limits for masses of clusters from the PSZ2 catalogue and the
future SRG/eROSITA survey. The mass limit for the PSZ2 catalogue is
adopted from \cite{PSZ2} and that for the SRG/eROSITA
survey has been calculated from its expected X-ray flux limit
$f_X>3\times 10^{-14}$~erg~s$^{-1}$~cm$^{-2}$ ($0.5$--$2$~keV), using
the mass -- X-ray luminosity relation from \cite{av09b}. Note that
for clusters with masses near $3\times 10^{14}~M_\odot$, a two times
lower mass detection threshold corresponds to approximately an order
of magnitude larger number of observed clusters.

We also note that the approximate lower mass limit for distant clusters
from our catalogue, $M_{500} > 3\times 10^{14}~M_\odot$, corresponds
to the masses of clusters that should be detected in the SRG/eROSITA
survey at all redshifts \citep[][]{chur15}. We thus conclude that
our catalogue contains a substantial fraction of the most massive
clusters that will be discovered by SRG/eROSITA (see also
Fig.~\ref{fig:m_z}). These clusters will be detected already during
the first year of the SRG survey and will be included in all
SRG/eROSITA cosmological samples and other large galaxy cluster
surveys in the Northern sky. 

We emphasize that we have found clusters near the detection limit
of the Planck survey. Therefore, a significant number of clusters 
can be missed in our catalogue near its supposed mass
threshold. Moreover, the angular resolution of the Planck telescope is in
fact insufficient for detection of distant clusters with masses
near $3\times 10^{14}~M_\odot$. For this reason, our catalog contains
a large number of projections (about 37\%), which are impossible to 
resolve with the Planck telescope. Therefore, the quality of
our catalogue is not high and it would be difficult to obtain any
cosmological constraints using this cluster sample. However, our
catalogue can be used to identify massive galaxy clusters that will
be discovered in future large cluster surveys, such as the planned
SRG/eROSITA all-sky X-ray survey.

\section{Conclusions}

We present a catalogue of galaxy clusters detected in Planck all-sky
Compton parameter maps and identified using data from the WISE and
SDSS surveys. Our catalogue comprises about 3000 clusters found in the
SDSS fields, with masses higher than approximately
$M_{500}\sim 3\times 10^{14}~M_\odot$. At redshifts above
$z\approx0.4$, the catalogue contains approximately an order of
magnitude more clusters than the 2nd Planck Catalogue of
Sunyaev-Zeldovich sources in the same fields of the sky. Therefore,
much more galaxy clusters can be discovered using the Planck survey
data, as compared to the number of clusters in the PSZ2 catalogue.
This is in agreement with the results obtained recently by
\cite{hurier17}, who found more SZ sources in Planck survey data with
the improved SZ sources detection algorithm.

In our work, clusters were found near the detection limit of the
Planck $y$-parameter map. Moreover, due to the insufficiently high
angular resolution of the Planck telescope, there is a large number of
projections in our catalogue (about 37\%). Therefore, the quality of
our catalogue is not high. However, this catalogue can be used to
identify massive galaxy clusters in large cluster surveys, such as the
upcoming SRG/eROSITA all-sky X-ray survey.

The majority of clusters in our catalogue already have spectroscopic
measurements of their redshifts. The number of the remaining clusters
(about 400) is not high, so that their redshifts can be measured at
ground telescopes equipped with efficient low-resolution 
spectrographs. Our group has been carrying out optical identifications
and redshift measurements for galaxy clusters from Planck surveys
using various ground telescopes
\citep{planckRTT150,PSZ1Addendum,Canary,vorobyev16}. To perform these 
observations, we have access to a significant amount of observing time
at the Russian-Turkish 1.5-m telescope, the 6-m telescope of SAO RAS (BTA),
and the 1.6-m Sayan Observatory telescope, which has recently been
equipped with a new medium and low resolution spectrograph 
\citep{azt33ik16}. We thus hope to measure all of the missing
spectroscopic redshifts in our cluster catalogue before the launch of
the SRG space observatory. 

\acknowledgements

The author is grateful to R.~Khatri for the detailed discussions of
his results and for very useful CO emission masks available for public
use. This work has been supported by Russian Science Foundation grant
14-22-00271.



\end{document}